\documentclass[prd,twocolumn,amsmath,amssymb,floatfix,superscriptaddress,nofootinbib]{revtex4-1}

\usepackage{graphicx}
\usepackage{amssymb}
\usepackage{amsmath}
\usepackage{bm}
\usepackage{color}

\DeclareMathAlphabet\mathbfcal{OMS}{cmsy}{b}{n}
\definecolor{darkgreen}{RGB}{50,150,0}
\definecolor{purple}{cmyk}{0.5,0.75,0,0}

\newcommand{\mr}{{m}}
\newcommand{\br}{{}}

\newcommand{\de}{Q}
\newcommand{\fake}{S}

\newcommand{\trueword}{real}
\newcommand{\Trueword}{Real}

\newcommand{\Comment}[1]{{}}

\definecolor{ultramarine}{rgb}{0.07, 0.04, 0.56}
\definecolor{cadmiumgreen}{rgb}{0.0, 0.42, 0.24}
\definecolor{indigo(dye)}{rgb}{0.0, 0.25, 0.42}
\usepackage[linktocpage=true]{hyperref}
\hypersetup{
colorlinks=true,
citecolor=ultramarine,
linkcolor=cadmiumgreen,
urlcolor=indigo(dye),
pdfauthor={},
pdftitle={},
pdfsubject={}
}

\begin{document}%%%%%%%%%%%%%%%%%%%%%%%%%%%%%%

\title{Separating the Universe into the Real and Fake}

\author{Wayne Hu}

\affiliation{Kavli Institute for Cosmological Physics, Department of Astronomy \& Astrophysics,  Enrico Fermi Institute, University of Chicago, Chicago, IL 60637}

\author{Chi-Ting Chiang}

\affiliation{C.N.~Yang Institute for Theoretical Physics, Department of Physics \& Astronomy,
Stony Brook University, Stony Brook, NY 11794}

\author{Yin Li}

\affiliation{Berkeley Center for Cosmological Physics, Department of Physics and Lawrence Berkeley National Laboratory,
University of California, Berkeley, CA 94720}
\affiliation{Kavli Institute for the Physics and Mathematics of the Universe (WPI),
UTIAS, 
The University of Tokyo, Chiba 277-8583, Japan}

\author{Marilena LoVerde}

\affiliation{C.N.~	Yang Institute for Theoretical Physics, Department of Physics \& Astronomy,
Stony Brook University, Stony Brook, NY 11794}

\begin{abstract}%%%%%%%%%%%%%%%%%%%%%%%%%%%%%%
The separate universe technique  provides a means of establishing  consistency relations 
between short wavelength observables and the long wavelength matter density fluctuations within which they evolve by absorbing the latter into the cosmological background.
We extend it to cases  where non-gravitational forces introduce a Jeans scale in
 other species like dynamical dark energy or massive neutrinos.
 The technique matches the
synchronous gauge matter density fluctuations to the local expansion using the acceleration equation and accounts for the temporal nonlocality
and  scale dependence of the long wavelength response of small scale matter observables, e.g.~the nonlinear power spectrum, halo abundance and the implied halo bias, and $N$-point correlation functions. 
 Above the Jeans scale, the local Friedmann equation relates the expansion to 
real energy densities and a curvature that is constant in comoving coordinates.   
Below
the Jeans scale, the curvature evolves and
acts like a fake density component.   In all cases, the matter evolution on small scales
is correctly modeled as we illustrate using scalar field dark energy with  adiabatic or isocurvature initial conditions across the Jeans scale set by 
its finite sound speed. 
\end{abstract}

\maketitle

\section{Introduction}%%%%%%%%%%%%%%%%%%%%%%%%%%%%%%

Describing the impact of a long-wavelength cosmological perturbation on 
small scale observables as a change in the background cosmology, or separate universe,
has proven very useful both conceptually and as a tool for making precise and consistent predictions between observables.
In the inflationary context, separate universe arguments provide insights into consistency relations
between the $N$-point functions \cite{Maldacena:2002vr, Creminelli:2004yq}, the evolution of
isocurvature fluctuations in multifield models \cite{Sasaki:1995aw}, and the observable
impact of compensated isocurvature fluctuations \cite{Grin:2011tf} within its domain of validity \cite{He:2015msa}.

 In the
late universe context, they have enabled studies of baryon acoustic oscillations \cite{SherwinZaldarriaga:12}, super sample power spectrum covariance \cite{TakadaHu:13,Lietal:14a},
 position dependent
power spectra \cite{Lietal:14b,Chiang:2014oga,Chiang:2015eza},
CMB lensing covariance \cite{Manzotti:2014wca}, and
dark matter halo bias 
\cite{Baldauf:2011bh,Baldauf:2015vio,Li:2015jsz,Lazeyras:2015lgp}. 
Moreover to the extent that the separate universe construction holds, these observable effects
can be modeled deep into the nonlinear regime with cosmological simulations
 \cite{Sirko:05,Gnedin:2011kj,Lietal:14b,Wagner:2014aka}, in principle compete with state-of-the-art treatments of astrophysical processes.
 In particular, long-wavelength modes have an impact on small scale observables that is
 nonlocal in time, complicating for example the modeling of the nonlinear power spectrum \cite{Ma:2006zk} and halo bias
 \cite{Hui:2007zh, Senatore:2014eva,LoVerde:2014pxa}.  In the separate universe approach these  can simply be modeled as a change in cosmological parameters.
 
The separate universe construction has traditionally been limited to large scales
where only gravitational forces act \cite{Dai:2015jaa}.   The effects of pressure or
anisotropic stress gradients in stabilizing fluctuations at the Jeans scale would seem to 
prevent replacing fluctuations with a separate homogeneous and isotropic universe.
While the non-relativistic matter is effectively pressureless on cosmological scales, the real
universe contains components that have relativistic stresses today or in the past,
e.g.~dark energy and massive neutrinos.   In these cases, the response of short
wavelength observables to long wavelength fluctuations can depend on their scale and redshift in addition to their amplitudes \cite{Parfrey:2010uy, LoVerde:2014pxa}.

In this work, we extend separate universe techniques to
the multi-component, relativistic case by introducing fictitious components below the Jeans scale that preserve
the illusion of a separate homogeneous and isotropic background from the perspective
of the small-scale matter distribution.
We begin in \S\ref{sec:SUconstruction} with the construction of a separate
expansion history that absorbs a long-wavelength matter density fluctuation even in this general case where non-gravitational forces act on other components.   Above the Jeans scale
set by these forces, the construction yields a separate universe with the real energy
densities and curvature of the long-wavelength fluctuations.   Below the Jeans scale,
non-conservation of curvature prevents these associations but retains the correct 
evolution of the small scale matter distribution. 

 In \S\ref{sec:components}, we discuss the relationship between the entropy,
 non-adiabatic stress, and curvature of the perturbations and the assignment of cosmological
 parameters in the separate universe above and  below the Jeans scale.   The
 separate universe construction determines how short wavelength
 observables respond to long wavelength modes including scale dependent and temporally nonlocal  effects as illustrated in \S\ref{sec:observable}.  We provide a concrete example
 of scalar field dark energy with adiabatic and isocurvature initial conditions in
 \S\ref{sec:scalar}.   We discuss these results in \S\ref{sec:discussion}.

\section{Separate Universe Construction}
\label{sec:SUconstruction}

In this section we develop the formal aspects of the separate universe construction in a fully
relativistic and multi-component context.
We begin in \S\ref{sec:separateexpansion} by defining the local expansion history that absorbs
a matter density fluctuation and its entire growth history into the background.   This construction corresponds to exactly matching the separate universe acceleration equation
with the synchronous gauge density perturbation and is always possible
 as shown in \S\ref{sec:Birkhoff}.  In 
\S\ref{sec:synchronous}, we find this construction also matches the Friedmann and energy conservation equations of the other components if the separate universe curvature implied by the curvature perturbation is constant (see also \cite{Dai:2015jaa}).
We relate this condition to the Jeans scales of these components in \S \ref{sec:Jeans}.

\subsection{Local Expansion and Density}
\label{sec:separateexpansion}
In the separate universe approach, we seek to absorb a long wavelength matter density fluctuation $\delta_\br =\delta\rho_\mr/\bar\rho_\mr$, including its entire growth history, into
the background of a separate universe \cite{Sirko:05,Baldauf:2011bh,Lietal:14a,Lietal:14b}
\begin{eqnarray}
\bar{\rho}_\mr(a) [1+\delta_\br(a)] &=&  \bar{\rho}_{mW}(a).
\end{eqnarray}
$``W"$ here and throughout 
denotes locally averaged or 
``windowed"  quantities on scales much smaller than the wavelength.
In terms of defining the local  cosmology, we can introduce the separate universe scale
factor through $\bar\rho_{\mr W} \propto a_W^{-3}$, which then defines the matter density parameters
\begin{eqnarray}
\frac{\Omega_\mr h^2}{a^3}(1+\delta_\br) &=& \frac{\Omega_{mW} h_W^{2}}{a_W^{3}}.
\label{eqn:equatedensity}
\end{eqnarray}
Here the Hubble constant $H_0=100 h$ km\, s$^{-1}$ Mpc$^{-1}$ and similarly  $H_{0W}$ is parameterized
by the dimensionless $h_W$.
Our convention is to set the scale factor of the separate universe $a_W$ to
agree with the global one $a$ at high redshift
\begin{equation}
\lim_{a\rightarrow 0} a_W(a)=a,
\end{equation}
where
\begin{equation}
\lim_{a\rightarrow 0} \delta_\br(a)=0.
\end{equation}
With this convention, the background energy density in the matter at the same numerical
values for $a$ and $a_W$ are always equal and hence
\begin{eqnarray}
\Omega_{mW} h_W^2 &=& \Omega_\mr h^2,
\label{eqn:omegamh2}
\end{eqnarray}
but the scale factors at the same time differ
\begin{eqnarray}
a_W  = \frac{a}{(1+\delta_\br)^{1/3}}&\approx& a\left(1- \frac{\delta_\br}{3} \right).
\label{eqn:aW}
\end{eqnarray}
In this construction, we assume that the two universes share a common universal
time. We shall see that common clocks of the two universe 
requires $\delta_\br$ to be specified in
synchronous gauge, a distinction that becomes important for scales near the horizon.  Equivalently, in
a gauge-invariant separate universe construction,  the
quantity of interest is the change in the $e$-folds
of the expansion \cite{Sasaki:1995aw}
which we can equate to the synchronous gauge 
 matter density perturbation since it evolves only via metric perturbations (see  \S \ref{sec:synchronous})
\begin{equation}
\delta N = \ln a_W-\ln a \approx -\frac{\delta_\br}{3}.
\end{equation}

The difference in scale factors also implies that the separate universe has 
a different expansion rate.
Using the definition  $H =\dot a/a$ and Eq.~(\ref{eqn:aW}), we obtain
\begin{equation}
\delta H^2 = H_W^2 - H^2 \approx -{\frac{2}{3} }H \dot\delta_\br = -\frac{2}{3} H^2 \delta_\br' .
\label{eqn:dh2}
\end{equation}
Here  overdots denote $d/dt$ in both the
global and separate universe whereas  $' = d/d\ln a$ will denote derivatives in the global
universe only.    

For the purpose of setting up the separate universe, 
it is sufficient to modify the expansion rate directly without making
further distinctions about its purported sources in the
 local Friedmann equation.
Notice that this construction in fact makes no direct use of  components
in the universe besides the matter.  
 Other components affect the construction  only through changing the matter growth
history $\delta_\br(a)$.
On the other hand this expansion history may or may not be generated by  the local Friedmann equation
with the physical components of the local universe.
To understand this issue, we consider the impact of the other components in the following sections.

\subsection{Birkhoff Theorem and Acceleration}

\label{sec:Birkhoff}

 In the Newtonian interpretation of the acceleration equation, 
the  non-relativistic matter can be considered as test particles 
 tracking the evolution of some region of physical radius $R$ in the  global universe.  It accommodates a
 perturbation $\delta_\br$ as long as $R$ is much smaller than its wavelength.
 
 Let us suppose that in addition to the matter there are additional density and pressure
 components 
 \begin{equation}
 \rho_Q=\sum_{J \ne \mr}\rho_J,\quad p_Q =\sum_{J\ne \mr} p_J.
 \end{equation} 
 The Birkhoff theorem relates the acceleration $\ddot R$ to the enclosed active gravitational mass
 \begin{equation}
\ddot R = -\frac{ 4\pi G}{3} [ \rho_\mr + \rho_\de + 3 p_\de ] R
 \end{equation}
and so employing the acceleration equation for the global universe
\begin{equation}
\frac{\ddot R}{R}   =\frac{\ddot a }{a}  - \frac{\Omega_\mr H_0^2}{2 a^3} \delta_\br 
- \frac{4\pi G}{3} (\delta\rho_\de + 3\delta p_\de).
\end{equation}
This radius can be absorbed into a separate universe scale factor if $a_W \propto R$.   
Using Eq.~(\ref{eqn:aW}) we also have
\begin{equation}
\frac{\ddot a_W}{a_W}   = \frac{\ddot a }{a} - \frac{2}{3} H \dot  \delta_\br - \frac{1}{3} \ddot \delta_\br.
\end{equation}
Thus the separate universe condition is
\begin{equation}
\frac{\ddot R}{R} -\frac{\ddot a_W}{a_W}=0
\end{equation}
or
 \begin{eqnarray}
\ddot  \delta_\br  + 2 H\dot  \delta_\br  &=&     \frac{3 \Omega_\mr H_0^2}{2 a^3} \delta_\br
  + 4\pi G  (\delta\rho_\de + 3\delta p_\de) \nonumber\\
&=& 4\pi G \sum_J (\delta\rho_J + 3\delta p_J)= 0.
\label{eqn:deltaBirkhoff}
  \end{eqnarray}
  We shall see in the next section that this is exactly the equation of motion for the synchronous gauge matter density perturbation. In the separate universe approach we are really
  just going to Lagrangian coordinates defined by the cold dark matter particles.   Only their
  relationship to Eulerian coordinates, quantified by $\delta_\br$, is influenced by other species in the universe. On scales that are well below
  the horizon, the distinction between synchronous gauge and other common gauges such as conformal Newtonian or comoving gauge becomes
  irrelevant and a Newtonian analysis for the density perturbation also applies.

\subsection{Synchronous Gauge and Friedmann Equation}
\label{sec:synchronous}

We can formalize the separate universe associations in the fully relativistic context of the Friedmann and acceleration equations.
The
assumption of a universal time implies that the $\delta_\br$ we absorb into the local 
background is the synchronous gauge fluctuation.  Its evolution must be compatible
with the local Friedmann and acceleration equations  for a \trueword\ separate universe construction.

In 
synchronous gauge the metric is given by $g_{00}=-1$, $g_{0i}=0$ and a perturbed
spatial metric
\begin{equation}
g_{ij} = a^2 (\gamma_{ij} + h_{ij}),
\end{equation}
where
$\gamma_{ij}$ is the 3-metric  of constant comoving curvature $K$ and the scalar metric perturbations for a mode of Laplacian wavenumber $k$ can be further decomposed into trace and tracefree pieces
\begin{equation}
h_{ij} = \frac{h_L}{3} \gamma_{ij} -  \left[ \nabla_i \nabla_j -\frac{1}{3}\gamma_{ij}\nabla^2\right] 
\frac{h_L+ 6\eta_T}{k^2}.
\end{equation}
Covariant differentiation and raising and lowering of spatial indices
is performed with respect to  $\gamma_{ij}$.  
In the separate universe $h_L$ performs the role of the perturbation to the scale factor $a$ and $\eta_T$ to
the spatial curvature $K$.   Specifically the perturbation to the 3D Ricci scalar  ${}^{(3)}R= 6 K/a^2$
on constant synchronous time slices (e.g.~\cite{Kodama:1985bj}),
\begin{equation}
 \delta K = -\frac{2}{3}(k^2-3K)\eta_T,
 \label{eqn:curvaturepert}
\end{equation}
whereas the  effect of $h_L$ is to change volumes and hence the effective Hubble rate 
by
\begin{equation}
\frac{\delta H}{H} = \frac{\dot h_L}{6 H}   = \frac{ h_L'}{6}.
\label{eqn:hubblepert}
\end{equation}
  Note that the effect of a single $k$-mode  perturbation is an
anisotropic change in the expansion rate.  For example, if $K=0$ the normal modes
are plane waves.  For a $k$-mode directed in the $x$ 
direction $e^{i kx}$, the scale factor and its time derivative
only change in the same direction $\delta \ln a_x = h_L/2$,
$\delta \ln a_y=\delta \ln a_z=0$.    Consequently, the separate universe construction only strictly
applies to the angle averaged response of local  observables to the long wavelength
mode, for example number densities of dark matter halos
 or angle averaged power spectra.

The $00$ and trace $ii$ Einstein equations relate the metric  to the
energy density $\delta\rho_J$ and pressure $\delta p_J$ perturbations  of the various components
(see e.g.~\cite{Kodama:1985bj} and \cite{Hu:2004xd,Gordon:2004ez} for a similar notation)
\begin{align}
 -\frac{k^2-3K}{(aH)^2} \eta_T + \frac{1}{2}  h_L' & = \frac{4\pi G}{H^2}  \sum_J \delta \rho_J,
 \label{eqn:etaT} \\
 h_{L}''  + \left( 2 + \frac{H'}{H} \right) h_{L}' &= -\frac{8\pi G}{H^2} \sum_J (\delta \rho_J + 3\delta p_J), 
 \label{eqn:hL}
\end{align}
which are themselves governed by the continuity and Navier-Stokes equation
 \begin{eqnarray}
 {\delta\rho_J'} + 3 ( {\delta\rho_{J}}+ \delta p_{J})
&=&
        - \frac{k \bar\rho_{J}}{aH} {u_{J}} - {{\bar\rho_{J}+\bar p_{J}} \over 2}  h'_{L} ,  
        \label{eqn:continuity}\\
      \bar\rho_{J} u_J' +( \bar\rho_{J} - 3 \bar p_J )u_J
&= &
\frac{k}{a H}\left[   { \delta p_{J} }-\frac{2}{3} (1-\frac{3K}{k^2}) p_J\pi_J \right] .
\nonumber
\end{eqnarray}

If the separate universe construction holds exactly, then these
 synchronous gauge  equations can be reabsorbed into the Friedmann, acceleration and
 energy conservation equations.
 We can already see  from the lack of a background Navier-Stokes equation that this
 can only be true if the momentum density 
$\bar\rho_J u_J$ generated by non-gravitational gradients in the isotropic stress $\delta p_J$ and anisotropic
stress $\pi_J$ can be ignored.   We  examine each of these equations in turn.

First let us check the  matter continuity equation (\ref{eqn:continuity}) and its  relation to
the perturbation to
the Hubble rate.  
For the matter $p_\mr=0$ and we further use the remaining gauge freedom of synchronous gauge to choose the freely falling observers to be on a grid of the pressureless matter particles themselves which sets $u_\mr=0$.  

With the shorthand convention $\delta=\delta_\mr$,  Eq.~(\ref{eqn:continuity}) becomes
\begin{equation}
\delta' = -\frac{1}{2} h_L' = -3 \frac{\delta H}{H}.
\label{eqn:deltah}
\end{equation}
This relation matches the separate universe construction in Eq.~(\ref{eqn:hubblepert}).

For the other components, if we take energy conservation in the background
\begin{equation}
\bar \rho_J' + 3 (\bar\rho_J+\bar p_J) = 0
\label{eqn:energy}
\end{equation}
and perturb the expansion rate we obtain the purely gravitational pieces of their
continuity equations (\ref{eqn:continuity}).   
The perturbation to the Hubble rate means that derivatives with respect to the
scale factor are perturbed as
\begin{eqnarray}
\frac{d}{d\ln a_W} &=& \frac{1}{H_W} \frac{d}{dt} = \frac{H}{H_W} \frac{1}{H} \frac{d}{dt} \nonumber\\
&\approx& \left( 1- \frac{\delta H}{H} \right) \frac{d}{d\ln a} ,
\label{eqn:primeW}
\end{eqnarray}
and so Eq.~(\ref{eqn:energy}) becomes for the perturbations
\begin{equation}
\delta \rho_J' + 3   (\delta \rho_J+\delta p_J) + 3 \frac{\delta H}{H} (\bar \rho_J+ \bar p_J)=0.
\label{eqn:superjeans}
\end{equation}
This matches the continuity equation (\ref{eqn:continuity}) when the effect of the divergence of the
non-gravitational
peculiar
velocities $u_J$ can be ignored.   These are generated through the Navier-Stokes
equation (\ref{eqn:continuity}) from pressure and anisotropic stress gradients which
give the condition
\begin{equation}
 \left( \frac{k}{a H} \right)^2 {\cal O} \left( \frac{\delta p_J}{\delta \rho_J},
\frac{ \bar p_J\pi_J}{\delta \rho_J}\right) \ll 1
\label{eqn:Jeans}
\end{equation} 
for them to change the energy density fluctuation negligibly.

Next through the change in the expansion rate (\ref{eqn:deltah}), the $ii$ Einstein equation (\ref{eqn:hL}) is related to the perturbation to the acceleration equation
\begin{equation}
H^2 + \frac{1}{2} \frac{d H^2}{d\ln a} = -\frac{4\pi G}{3} \sum_J (\bar\rho_J+ 3\bar p_J).
\label{eqn:acceleration}
\end{equation}
Converting derivatives of the scale factor with Eq.~(\ref{eqn:primeW}),
we obtain the perturbed acceleration equation in terms of the global $a$ as
\begin{equation}
\left( \frac{\delta H}{H} \right)'  + \left( 2 + \frac{H'}{H} \right) \frac{\delta H}{H} = 
-\frac{4\pi G}{3H^2}\sum_J (\delta\rho_J+ 3\delta p_J),
\end{equation}
which matches Eq.~(\ref{eqn:hL}) given (\ref{eqn:deltah}).  Notice that unlike  the
continuity equations, the acceleration equations  for the background and perturbations
take exactly the same form with no  further restrictions on scales, in agreement with
the discussion of the Birkhoff theorem in the previous section. 

Finally  the $00$ Einstein equation~(\ref{eqn:etaT}) is the perturbation to the Friedmann equation
\begin{equation}
H^2 + \frac{K}{a^2} = \sum_J \frac{8\pi G}{3} \bar\rho_J,
\end{equation}
or
\begin{equation}
\delta H^2 + \frac{\delta K}{a^2} = \sum_J \frac{8\pi G}{3} \delta \rho_J,
\label{eqn:dFriedmann}
\end{equation}
with the associations of Eq.~(\ref{eqn:curvaturepert}) and (\ref{eqn:hubblepert}).
Conversely, these perturbations can be reabsorbed into a Friedmann equation of a local,
 separate universe if the curvature fluctuation
  $\delta K(a)$ can be replaced by a new constant 
 curvature $K_W$ in  coordinates that comove with $a_W$
 \begin{equation}
 K_W \equiv \frac{a_W^2}{a^2} ( K + \delta K) \approx K+ \delta K - \frac{2}{3}K \delta.
 \label{eqn:KW}
 \end{equation}
If the global universe is flat $K=0$, the curvature perturbation
 $\delta K$ itself must be constant.   Thus the existence of a \trueword\ separate universe is intimately related to the conservation of curvature perturbations outside
 the horizon \cite{Bardeen:1980kt}.
If $K\ne 0$, the curvature perturbation must
 evolve to account for the different local scale factor of the perturbed universe.
 
 Thus from the perspective of matching the synchronous gauge perturbation equations to
 background equations in the separate universe, the continuity equation requires non-gravitational
 flows $u_J$ to be negligible and the Friedmann equation requires the 
 curvature $K_W$ to be constant.   We shall now see that these are essentially the same criteria.

 \subsection{Curvature Conservation and  Jeans Scale}
 \label{sec:Jeans}
 
 To better understand why the constancy of curvature is related to having negligible non-gravitational
 flows, 
 it is useful to examine the redundant $0i$ Einstein equation
 which directly gives the evolution equation for the curvature fluctuation
 \begin{equation}
 (1-\frac{3K}{k^2})\eta_T' - \frac{K}{2k^2} h_L' = \frac{4\pi G}{H^2} \frac{ aH}{k}\sum_J \bar\rho_J u_J.
 \label{eqn:curvevol}
 \end{equation}
 This equation has no equivalent in the background given homogeneity and isotropy and
 so in the separate universe construction should produce a tautology $0=0$.
 Using Eq.~(\ref{eqn:curvaturepert}) and (\ref{eqn:KW}) we can rewrite Eq.~(\ref{eqn:curvevol}) as
 \begin{equation}
 K_W' = -\frac{8\pi G}{3} \frac{k^2}{H^2}  \frac{ aH}{k}\sum_J \bar\rho_J u_J.
\end{equation}
The curvature is effectively constant when we can ignore the non-gravitational velocities $u_J$.
 Given the Navier-Stokes equation (\ref{eqn:continuity})
\begin{equation}
\bar\rho_J u_J = \frac{k}{aH} {\cal O} (\delta p_J, p_J\pi_J),
\label{eqn:momentumjeans}
\end{equation}
we obtain the estimate
\begin{equation}
 K_W'  = { 8\pi G}\frac{k^2}{H^2}  {\cal O}(\delta p_T, p_T\pi_T),
 \end{equation}
 where ``$T$" denotes the total of all components.
 We call the scale $k_T$ at which this change in curvature per efold becomes comparable to the
 curvature fluctuation itself,
\begin{equation}
\frac{K_W'}{\delta K}  = {\cal O}(1),
\end{equation}
the total Jeans scale.

 While this defines the total Jeans scale and relates it to non-gravitational flows, it is useful to estimate its value in particular cases to 
 relate it to more conventional definitions. 
 For metric perturbations sourced by growing total density fluctuations $\delta\rho_T$, where 
$|\delta p_T/\delta\rho_T| \lesssim 1$, $h' = {\cal O}(4 \pi G \delta\rho_T/H^2)$.   Then
 Eq.~(\ref{eqn:etaT})  gives the order of magnitude of the curvature fluctuation itself
 \begin{equation}
 \delta K = 4\pi G a^2 {\cal O}(\delta \rho_T) 
 \label{eqn:Korder}
 \end{equation}
and so \begin{equation}
\frac{K_W'}{\delta K} = \left( \frac{k}{a H} \right)^2 {\cal O} \left( \frac{\delta p_T}{\delta \rho_T},
\frac{\bar p_T\pi_T}{\delta \rho_T}\right).
\label{eqn:JeansT}
\end{equation}
For a  single component with only isotropic, adiabatic stresses, 
$\delta p_T/\delta \rho_T = p_T'/\rho_T' \equiv c_{Ta}^2$, this scale corresponds to the usual Jeans condition that pressure prevents further growth below  the sound horizon or Jeans scale
$c_T k_T/aH \approx 1$.  If the sound speed is subluminal $c_T < 1$, the Jeans scale is always below the horizon scale.   Note that in this case there is no difference between the constant curvature
condition (\ref{eqn:JeansT}) and the negligible non-gravitational flows condition (\ref{eqn:Jeans}).

More generally, the total pressure is composed of the adiabatic $(p_J'/\rho_J')\delta\rho_J=c_{Ja}^2\delta\rho_J$,
internal non-adiabatic stress $\Gamma_J$ of the various components,
 \begin{equation}
\delta p_J =  c_{Ja}^2 \delta \rho_J + p_J\Gamma_J,
 \end{equation}
and their relative entropy fluctuations with the matter
\begin{align}
S_{J \mr} &= \frac{\delta \rho_J }{\bar\rho_J +\bar p_J } - \frac{\delta \rho_\mr}{\bar\rho_\mr+\bar p_\mr} ,
\label{eqn:entropy}
\end{align}
such that
\begin{equation}
\delta p_T = c_{Ta}^2\delta \rho_T + p_T\Gamma_T,
\end{equation}
with the total non-adiabatic stress
  \begin{equation}
 p_T\Gamma_T 
 = \sum_J \left[ p_J \Gamma_J +  S_{J\mr}(\rho_J+p_J)( c_{Ja}^2 - c_{Ta}^2 ) \right].
 \end{equation}
 In the general case, the adiabatic sound speed $c_{Ta}$ no longer bounds the total
 pressure.   
 
  Entropy fluctuations allow initial isocurvature conditions where the 
   total Jeans scale can be made arbitrarily large compared with the horizon.  In this case we need to slightly generalize the estimate  (\ref{eqn:JeansT})
 since the total pressure fluctuation can be larger than the total density perturbation and we need
 to separate their contribution through  $h_L'$  to $\eta_T$ in Eq.~(\ref{eqn:etaT}).
This division can be readily identified by using the final synchronous Einstein equation, the redundant
$ij$ tracefree equation
\begin{eqnarray}
&&-\left(\frac{k}{aH}\right)^2 \eta_T + \frac{  h_L''+6\eta_T''}{2} + \left(3 + \frac{H'}{H}\right) 
\frac{  h_L'+6\eta_T'}{2}
\nonumber\\
&&\qquad
 =  -\frac{ 8\pi G}{H^2} \sum_J p_J \pi_J,
\end{eqnarray}
and combining Eq.~(\ref{eqn:etaT}) and  (\ref{eqn:curvevol}) into
 \begin{equation}
 \left(1-\frac{3K}{k^2} \right)\left[  \frac{h_L'+6\eta_T'}{2} -\frac{k^2}{(aH)^2} \eta_T  \right]= 
 \frac{4\pi G}{H^2} \delta \rho_{Tc},
 \end{equation}
  where
 \begin{equation}
\delta \rho_{Tc}=  \sum_J \left[ \delta\rho_{J}
+ 3\left(\frac{aH}{k}\right) \rho_J u_J \right]
\end{equation}
 defines the density perturbation in comoving gauge \cite{Bardeen:1980kt}.   With the assumption that
 the total anisotropic stress is negligible outside the horizon 
  \begin{equation}
\delta K = {4\pi G}a^2  {\cal O}( \delta \rho_{Tc} ),
\end{equation}
which generalizes Eq.~(\ref{eqn:Korder}) for total density fluctuations that grow from isocurvature initial conditions.

Thus isocurvature conditions result when the contributions
to the comoving gauge density perturbations cancel between species of different
sound speeds leaving finite pressure perturbations.  
 In this case, the  curvature fluctuation is initially small but evolves significantly
so that its small impact on the local curvature cannot be captured as a \trueword\
separate universe.  On the other hand, since our universe possesses  adiabatic
or initial curvature fluctuations, even if isocurvature modes $S_{J\mr}$ are comparable to
the curvature fluctuations, their impact on the separate universe curvature above the horizon is
negligible (see \S\ref{sec:scalar}).   We shall also see there that an entropy fluctuation
$S_{J\mr}$ forms dynamically from initial curvature fluctuations if $J$ has an intrinsic non-adiabatic stress $\Gamma_J$ as it must for a dynamical dark energy component (see \S \ref{sec:scalar} and \cite{Hu:1998kj}).

Finally, for collisionless particles, free streaming generates anisotropic
stress and their gradients generate higher moments.   Anisotropic stress gradients
also act as an effective viscosity in the Navier-Stokes equation (\ref{eqn:continuity})
generating $u_J$ and setting an effective Jeans scale in Eq.~(\ref{eqn:momentumjeans}) called
the free streaming scale.  

The distinction between the constant curvature condition (\ref{eqn:JeansT}) and the non-gravitational flows
condition  Eq.~(\ref{eqn:Jeans}) is that the latter sets a Jeans scale for each component.
If a component
 contributes negligibly to the total then  it has a negligible effect on the curvature even
 below its Jeans scale.   In \S\ref{sec:scalar}, we shall see an example where in the
matter dominated regime, the dark energy has its own Jeans scale that is much larger
than the total Jeans scale but does not impact the matter evolution.    On the other
hand this distinction becomes irrelevant if, as in this case, the component eventually does
dominate the expansion.

In summary, 
if the wavelength of $\delta_\br$ is larger than the Jeans scales of all components
we call its absorption into a local background
as  a ``\trueword" separate universe construction since all of the perturbation equations 
can be absorbed into the background with real energy density and curvature
in the Friedmann equation.   If the wavelength is shorter than the total Jeans scale, we
call this a ``fake" separate universe construction.   
In this case from the perspective of the matter, the local universe obeys an effective
Friedmann equation.    Here the curvature in comoving coordinates evolves but 
as we shall see in the next section it be considered as an effective energy density component
for the matter dynamics.  

This correspondence enables and justifies 
 a separate universe treatment of the response
 of  small scale cosmological observables to a long wavelength density perturbation
 even if that wavelength is below the total
 Jeans scale of the system.

\section{Separate Universe Components}
\label{sec:components}

In the previous section, we have shown that a long wavelength matter density fluctuation
can always be reabsorbed into the background with an appropriate adjustment of the
expansion rate to its local or separate universe value.   By construction, this approach satisfies
the acceleration equation or the Birkhoff theorem exactly.  For the separate universe
Friedmann equation to be truly satisfied in terms of real energy densities and curvature, 
the wavelength must be  much longer than the total Jeans scale in order for the curvature
to be constant in comoving coordinates.    

Below the Jeans scale, 
if the non-matter components only influence the small scale matter evolution through the expansion
rate,  they can be described by an effective
energy density component.   In this ``fake" 
separate universe, the dynamical impact
of a changing curvature is assigned to this fictitious energy density component.

In this section, by matching  parameters in the global and separate universe Friedmann
equation, we establish this correspondence explicitly.  We begin in \S\ref{sec:matching} with the
comparison of the two Friedmann equations.    We relate parameters
in the \trueword\ separate universe  in \S\ref{sec:true} and in the fake separate universe
in \S\ref{sec:fake}.

\subsection{Friedmann Matching}
 \label{sec:matching}

The Friedmann equation in the global background universe can without loss of generality
be written as
 \begin{eqnarray}
\frac{H^2}{H_0^2}  &=& \frac{\Omega_{\mr }}{a^3} + \Omega_{\de}F_\de(a)
+ \frac{ \Omega_{K }}{a^2} ,
\label{eqn:Friedmannglobal}
\end{eqnarray}
where  $\de$ represents the sum over all components aside from the pressureless matter.  
Here $F_\de(a=1)=1$ and so defines  $\bar\rho_\de(a)$ relative to its value today.  Its derivative
gives the equation of state parameter
\begin{equation}
 \frac{d \ln F_\de}{d \ln a} = -3(1+w_\de).
 \end{equation}
 Let us first try a naive method for absorbing the energy density fluctuations 
 $\delta_{\mr}$ and $\delta_\de$ into a local background.   Taking an equation of the same form
  \begin{eqnarray}
{H_W^2}  &=&{H_{0W}^2} \left[ \frac{\Omega_{\mr W}}{a_W^3} + \Omega_{\de W}F_\de(a_W)
+ \frac{ \Omega_{KW }}{a_W^2}\right] , \label{eqn:Friedmannad}
\end{eqnarray}
we can attempt to set the local parameters using the Friedmann equation with perturbed energy densities
\begin{eqnarray}
{H_W^2}  
  &=&{H_{0}^2} \left[ (1+\delta_\br)\frac{\Omega_{\mr}}{a^3} +(1+\delta_\de) \Omega_{\de}F_\de(a)\right]
  \nonumber\\
&& +{H_{0W}^2}   \frac{ \Omega_{KW }}{a_W^2}.\label{eqn:Friedmannguess}
\end{eqnarray}
 Using the $a_W(a)$ relationship (\ref{eqn:aW}), the Hubble rates in Eq.~(\ref{eqn:Friedmannad}) and Eq.~(\ref{eqn:Friedmannguess}) coincide if
 \begin{eqnarray}
 \Omega_{\mr W}H_{0W}^2 &=& \Omega_{\mr} H_0^2 , \nonumber\\
  \Omega_{\de W}H_{0W}^2 &=&  \Omega_{\de} H_0^2, \nonumber\\
  \delta_\de &=& -\frac{1}{3} \frac{d \ln F_\de}{d\ln a} \delta_\br,
  \label{eqn:naiveSU}
  \end{eqnarray}
i.e.\ if the energy densities agree when $a$ and $a_W$ have the same numerical value and the entropy perturbation (\ref{eqn:entropy}) 
\begin{align}
S_{\de \mr} 
&= \frac{\delta_\de }{1+w_\de } - \delta_\br 
\end{align}
between the components vanishes.
In this case the same shift in the scale factor of Eq.~(\ref{eqn:aW}) that absorbs the matter
fluctuation would absorb the $\de$ fluctuation  as well for all time.   However  Eq.~(\ref{eqn:naiveSU}) is not a necessary condition 
and in fact cannot be stably satisfied if $\de$ contains dynamical dark energy components
(see \S \ref{sec:scalar} and \cite{Hu:1998kj}).

More generally in the separate universe construction, the synchronous gauge matter density fluctuation 
$\delta$ defines the Hubble rate in the separate universe $H_W$ through Eq.~(\ref{eqn:dh2})
and the relationship between the scale factors at constant time through Eq.~(\ref{eqn:aW}).
We can relate $H_W^2$ to the energy densities and curvature to a more general form of
the Friedmann equation
\begin{eqnarray}
\frac{H_W^2}{ H_{0W}^2  } &=&  \frac{\Omega_{\mr W}}{a_W^3} + \Omega_{\de W}F_\de(a_W)
+ \frac{ \Omega_{K W}}{a_W^2} 
\nonumber\\&&
   + \Omega_{\fake W} F_\fake(a_W)  ,
   \label{eqn:Friedmannlocal}
   \end{eqnarray}
where $F_\fake(a_W=1)=1$. 
The introduction of the $\fake$ component allows us to match any expansion history, not just those defined by perturbations to the $\Omega_\de$ of the global universe.   We shall see next that its presence
indicates an entropy perturbation or non-adiabatic stress for a \trueword\ separate universe and
a fictitious energy density that accounts for the evolution of the curvature in a fake separate universe.

We can obtain these correspondences by
equating the difference in the Friedmann equation Hubble rates
defined by Eq.~(\ref{eqn:Friedmannlocal}) and (\ref{eqn:Friedmannglobal}) 
 to that required by Eq.~(\ref{eqn:dh2}) to match the acceleration equation.  Keeping terms linear in $\delta_\br$
\begin{eqnarray}
\frac{\delta H^2}{H_0^2} &\equiv&  -{\frac{2}{3} } \frac{H^2}{H_0^2 } \delta_\br' \nonumber\\
&=&
\frac{\Omega_\mr}{a^3} \delta_\br - \frac{\Omega_\de}{3}  F_\de' \delta_\br + \frac{2}{3} \frac{\Omega_K}{a^2}  \delta_\br +
\frac{2 \delta h}{h} \frac{1}{a^2}  \nonumber\\
&&+ \Omega_{\fake W}\left[ F_\fake -\frac{1}{a^2} \right].
\label{eqn:matching}
\end{eqnarray}
Here $2 \delta h/h \approx (H_{0W}^2-H_0^2)/ H_0^2$ is a constant associated
with the expansion rates at two different times but the same numerical value of the scale factor, namely $a_W=1$ and $a=1$.   Here and below we use the notation $\delta X = X_W-X$ for a parameter 
 $X$.
  Note that $\Omega_{\fake W} \propto \delta_\br$ since this component is absent
in the global universe. 
 
In particular $\delta h/h$ is defined by evaluating Eq.~(\ref{eqn:matching}) at $a=1$ using
$\delta'(a=1) \equiv \delta_0'$
\begin{equation}
2 \frac{\delta h}{h}  =  -{\frac{2}{3} }\delta_0'  -{\Omega_\mr} \delta_{\br 0} +\frac{ F_\de'}{3}   \Omega_\de \delta_{\br 0} - \frac{2}{3} {\Omega_K}  \delta_{\br 0}.
\label{eqn:dh}
\end{equation} 
Equality of the physical energy densities at the same scale factor sets
\begin{eqnarray}
\frac{\delta \Omega_m}{\Omega_m}
=\frac{\delta \Omega_\de}{\Omega_\de}
=
{- 2 \frac{\delta h}{h}},
\end{eqnarray}
and
\begin{equation}
 \Omega_{\fake W} +\Omega_{KW}  = 1-\Omega_{\mr W}-\Omega_{\de W},
 \label{eqn:sumomega}
 \end{equation}
 by definition of $H_{0W}^2$. 
While this assumption for $\delta\Omega_{\de}$ is not fully general, 
 we can absorb any remaining difference  into the $\fake$ component.   In other words, we
 take $\delta\Omega_{\de}$ to define the division into $\de$ and $\fake$ components in the separate universe.  
  We shall make explicit use
 of this fact in the dark energy isocurvature example in \S \ref{sec:scalar}.
 
These relations set the separate universe  parameters of the energy density components that exist in the global
universe.  If we take $\Omega_{\fake W}=0$, then the curvature is also determined and there is
no additional freedom that can be used to satisfy Eq.~(\ref{eqn:matching}) at $a< 1$.
Thus for a general evolution of $\delta_\br$, $\Omega_{\fake W} \ne 0$ is required, and we
can define $F_\fake$ so as to satisfy Eq.~(\ref{eqn:matching}).   Conversely for any 
desired evolution of $\delta_\br$, we can always construct a well defined expansion history
using $\fake$ to satisfy both the acceleration and Friedmann equations.

\subsection{\Trueword\ Separate Universe}
\label{sec:true}

The distinction between a \trueword\ and fake separate universe construction depends on
whether the Friedmann components $\Omega_{\fake W}$
and $\Omega_{KW}$ truly represent an energy density 
and curvature in the local universe.    In the construction of Eq.~(\ref{eqn:sumomega}),  only their sum and not their individual values
are specified.   This ambiguity is related to the fact that in the Friedmann and acceleration equations  it is not
possible to distinguish between an energy density that scales as $1/a^2$ and a curvature 
component.    

However, curvature has geometric effects which distinguish it and 
moreover we can relate the curvature
perturbation and the separate universe curvature using Eq.~(\ref{eqn:KW}) 
 at $a_W=1$, 
\begin{eqnarray}
\Omega_{K W} &\equiv &-\frac{K_W}{H_{0W}^2} \nonumber\\
& =& -\frac{K}{H_0^2}-\frac{\delta K}{H_0^2} +
\frac{K}{H_0^2}\left( \frac{2}{3} \delta + 2 \frac{\delta h}{h}\right). \\
\frac{\delta K}{H_0^2} &=& \frac{8\pi G}{3 H_0^2} \sum_J \delta \rho_J + \frac{2}{3} \delta_\br',
\end{eqnarray}
where we have used Eqs.~(\ref{eqn:curvaturepert}) and (\ref{eqn:etaT}).
Combining these equations with Eq.~(\ref{eqn:dh}), we obtain
 \begin{eqnarray}
\delta \Omega_{K} & =&  -\Omega_\de \delta_{\de } - \frac{F_\de'}{3} {\Omega_\de}\delta_{\br }
+ 2 (1-\Omega_K) \frac{\delta h}{h} \nonumber\\
&=& -\Omega_\de (1+ w_\de) S_{\de \mr} + 2 (1-\Omega_K) \frac{\delta h}{h} .
\label{eqn:deltaK}
\end{eqnarray}
Finally using Eq.~(\ref{eqn:sumomega}), we obtain
\begin{equation}
\Omega_{\fake W} = \Omega_\de (1+ w_\de) S_{\de \mr}
\label{eqn:Omegafake}
\end{equation}
so that this component is associated with the entropy perturbation.  All
of the above relationships for separate universe cosmological parameters $\Omega_{J W}$ in terms of
global universe perturbations are assumed to
be evaluated at $a_W=1$ and note that $\delta({a_W}=1) \approx \delta(a=1)\equiv \delta_0$.

We obtain the same criteria from the standpoint of absorbing the energy density associated
with $\delta_\de$
into the background at an arbitrary $a_W(a)$
\begin{eqnarray}
&& H_0^2 \Omega_\de F_\de(a) (1+\delta_\de) \nonumber\\
&&\quad =  H_{0W}^2 \Omega_{\de W} F_\de(a_W) %\nonumber\\&&
+ H_{0W}^2 \Omega_{\fake W} F_\fake(a_W)  
\nonumber\\
&& \quad \approx  H_0^2 \Omega_{\de } \left[F_\de(a) - \frac{F_\de'(a)}{3} \delta_\br \right]
+ H_{0}^2 \Omega_{\fake W } F_\fake(a)  .
\end{eqnarray}
Employing the definition of the entropy 
\begin{equation}
F_\de \delta_\de = -\frac{F_\de'}{3} (S_{\de \mr} + \delta_\br),
\end{equation}  
we infer
\begin{equation}
\Omega_{\fake W} F_\fake(a) = -\Omega_\de \frac{F_\de'(a)}{3} S_{\de m}(a) ,
\label{eqn:fstrue}
\end{equation}
which gives Eq.~(\ref{eqn:Omegafake}) at $a=1$ and defines the energy density scaling in terms of
the evolution of the entropy.   Note that the presence of an evolving entropy fluctuation does not
prevent a \trueword\ separate universe matching: each component represents a real energy
density that exists in the global universe.   It simply means that in the separate universe,
the background energy density components do not obey the same equations of state as in the
global universe.
On the other hand we shall see next that non-conservation of the separate universe curvature
below the Jeans scale does indicate that the separate universe construction involves
fake components.

\subsection{Fake Separate Universe}
\label{sec:fake}

While the remapping of perturbations onto  separate universe cosmological parameters in the previous
section may seem fully general, it implicitly assumes that the separate universe curvature
$K_W = $\,const.\ whereas it is actually constructed in 
Eq.~(\ref{eqn:KW}) out of the dynamical curvature and scale factor fluctuations in the global universe
\begin{equation}
K_W' =  \delta K' - \frac{2}{3} K \delta_\br'. 
\end{equation}
 Setting $K_W'=0$ and combining the synchronous gauge metric
equations, we obtain the condition
\begin{align}
(\delta \rho_\de)' + 3(\delta \rho_\de+\delta p_\de) = 
        (\bar\rho_{\de}+\bar p_{\de})\delta_\br'\,.
\end{align}
Not surprisingly this is exactly the same condition in Eq.~(\ref{eqn:superjeans}) for which
the continuity equation Eq.~(\ref{eqn:continuity}) can be written as a perturbation to
the background
energy conservation equation.   This condition holds to good approximation  for scales above the Jeans scale for $\de$ including any evolution in $S_{\de \mr}$.   In this case $\delta_\de$ evolves as
in a separate universe.  

Below the Jeans scale, the matter fluctuations still behave in a way that can be absorbed
into a separate universe expansion rate but one that does not obey a true Friedmann equation.
If we
use the curvature and entropy decomposition in Eq.~(\ref{eqn:deltaK}) and (\ref{eqn:Omegafake}), we must 
 allow the curvature contribution to the
Friedmann equation to have a general evolution
\begin{eqnarray}
\frac{H_W^2}{ H_{0W}^2} &=& \frac{\Omega_{\mr W}}{a_W^3} + \Omega_{\de W}F_\de(a_W)
+ { \Omega_{K W}}F_K(a_W)\nonumber\\
&&
   + \Omega_{\fake W} F_\fake(a_W) ,
   \end{eqnarray}
where 
\begin{equation}
\Omega_{K W} F_K(a_W) \equiv  - \frac{K_W(a_W)}{H_{0W}^2 a_W^2}.
\end{equation}
   In this case  $K_W$ is a  non-constant background curvature which has a ``fake"
   equation of state $w_{KW}\ne -1/3$.

 Alternatively, we can combine the curvature fluctuation  into the effective energy density
$\fake$,
\begin{eqnarray}
\delta \Omega_K&=& 0 ,\nonumber\\
\Omega_{\fake W} &=& -\delta\Omega_{\mr} - \delta\Omega_\de,
\end{eqnarray}
which evolves according to
\begin{eqnarray}
 \Omega_{\fake W} F_\fake(a) &=&\frac{\Omega_{\fake W}}{a^2}  -{\frac{2}{3} }  \frac{H^2 }{H_0^2 } \delta_\br' -
\frac{\Omega_\mr}{a^3} \delta_\br+  
\frac{F_\de'  }{3}  \Omega_\de\delta_{\br} \nonumber\\
&&
 - \frac{2}{3} \frac{\Omega_K}{a^2}  \delta_\br -
\frac{2 \delta h}{h} \frac{1}{a^2} .
\end{eqnarray}
In this case, $\fake$ is a fake energy density component that accounts for both the
curvature and entropy fluctuations.
Of course in practice given that either alternative involves a fake component to the Friedmann equation 
one can also simply set an expansion rate $H_W$ without dividing its
sources into separate energy density and curvature components.  

In the intermediate regime, where the growth of $\delta_\br$ depends on scale, we can
either analyze the separate universes for each $k$-mode in turn or construct the
real space $\delta_\br$ that corresponds to the sum over the modes that contribute
to the local mean averaged over a given physical scale.

\section{Observable Response}
\label{sec:observable}

In both the \trueword\ and fake or super and sub Jeans scale separate universe constructions, the matter distribution on small scales responds to a long wavelength fluctuation  as if it were in a separate
universe, i.e.\  through the change in cosmological parameters associated with the separate universe.  In general, this means that these observables 
respond not just to the change in the local mean density at the epoch that they are observed
but also the whole history of
its evolution.   If this growth history depends on the scale of the long wavelength mode
then the observable response will as well.   

For nonlinear observables such as the bias and abundance of dark matter halos or 
the nonlinear matter or halo power spectra, we can employ cosmological simulations in the
separate universe
 \cite{Sirko:05,Gnedin:2011kj,Lietal:14b,Wagner:2014aka} to calibrate this response \cite{Li:2015jsz,Lietal:14b,Lietal:14a,Baldauf:2015vio,Li:2015jsz,Lazeyras:2015lgp}.   In future work we will present results 
from cosmological simulations in the \trueword\ and fake separate universe for dynamical dark energy
and massive neutrino models.

Here we illustrate the ideas with the observable being the growth of structure or power spectrum in the linear
regime.  If the short wavelength mode is much smaller than the
Jeans length of the other components $\de$, the evolution of $\delta_{W}(a)$ obeys the usual growth
equation obtained by setting the $J=\de$ components to zero in 
Eq.~(\ref{eqn:hL}), but with the scale factor and expansion rate of the separate universe
\begin{equation}
\frac{d^2 \delta_W}{d \ln a_W^2}  + \left(2 + \frac{d \ln H_W}{d\ln a_W} \right) \frac{d \delta_W}{d \ln a_W} 
= %\nonumber\\&&\quad 
\frac{3}{2}\frac{H_{0W}^2}{H_W^2}  \frac{\Omega_{\mr W}}{a_W^{3}}
\delta_W .
\end{equation}
Note that here the density fluctuation and growth is relative to the separate universe
mean $\rho_{\mr W}=\rho_{\mr}(1+\delta_\br)$.  
Using the relationship between the separate and global universes $(\ref{eqn:aW})$,
$(\ref{eqn:dh2})$  and $(\ref{eqn:primeW})$, we can rewrite 
this in the global coordinates as 
\begin{equation} 
\delta_W'' + \left( 2 -\frac{2}{3} \delta_\br' + \frac{d\ln H}{d\ln a} \right) \delta_W' =
\frac{3}{2}\frac{H_{0}^2}{H^2}  \frac{\Omega_{\mr}}{a^{3}}(1+\delta_\br)
\delta_W .
\label{eqn:dWgrowth} 
\end{equation}
Since the change in the growth due to $\delta_\br$ is itself small, we  can
expand
\begin{equation}
\delta_W = \delta_- + \epsilon,
\end{equation}
where $\delta_-$ is given by the unperturbed sub Jeans scale growth, i.e.~by setting $\delta_\br=0$
in Eq.~(\ref{eqn:dWgrowth}).  Here
$\epsilon =  {\cal O}( \delta_\br)\delta_-$ is the second order correction from the long-wavelength mode
 that obeys
\begin{eqnarray}
\epsilon'' + \left(2 + \frac{H'}{H} \right) \epsilon ' -  \frac{3}{2}\frac{H_0^2}{H^2} \frac{ \Omega_\mr }{a^{3}} 
 \epsilon   \nonumber\\
= \frac{2}{3} \delta_\br' \delta_-' + \frac{3}{2} \frac{H_0^2}{H^2} \frac{ \Omega_\mr}{ a^{3}} 
\delta_\br \delta_-.
\end{eqnarray}
Notice that both the long wavelength perturbation to the scale factor $\delta_\br$ and the perturbation to the Hubble rate $\delta_\br'$ enter as sources.

Solving this system, we can define the growth response function
\begin{equation}
\frac{d\ln D_W}{d\delta} = \frac{\epsilon}{\delta_{\br} \delta_-}.
\label{eqn:DWresponse}
\end{equation}
For growing modes in the matter dominated limit,
\begin{equation}
\frac{d\ln D_W}{d\delta} = \frac{13}{21} ,
\end{equation}
which is the usual second order result \cite{Bernardeauetal:02}, accounting for the difference due to
fluctuations being measured with respect to $\rho_{\mr W}$, i.e.~in the global universe
\begin{equation}
\frac{d\ln D}{d\delta} = \frac{d\ln D_W}{d\delta}+ 1.
\end{equation}
This implies that the power spectrum response to $\delta$ 
in
the linear regime \cite{Lietal:14a}
\begin{equation}
\frac{\partial \ln P}{\partial \delta} = 2\frac{d\ln D}{d\delta}  -
\frac{1}{3} \frac{d\ln k^3 P}{d\ln k }
\end{equation}
depends on the growth history of the long wavelength mode $\delta_\br$ which can itself depend on scale.   The second term on the RHS comes from the dilation of scales due to the
separate universe scale factor \cite{Lietal:14a}.
   One consequence of this is that the squeezed bispectrum and trispectrum becomes dependent on 
the scale of the long-wavelength mode.   The latter also causes the super-sample covariance of the power spectrum \cite{Lietal:14a}
to also depend on which modes contribute to the local mean within the sample.

These relations are also useful for setting up cosmological simulations of the separate 
universe with the same initial conditions as the global universe given a power spectrum normalization at $a_W=1$ commonly used in codes.

\section{Scalar Field Dark Energy}
\label{sec:scalar}

As an illustration of the concepts in the previous sections, let us consider the concrete
example of scalar field dark energy with a Lagrangian  \cite{ArmendarizPicon:2000dh}
\begin{equation}
{\cal L}=P(X,Q), \quad 
X=-\frac{1}{2}\nabla^\mu Q \nabla_\mu Q.
\end{equation}
In this case $\de$ represents a single
component, the scalar field itself, and takes the form of a perfect fluid with no anisotropic stress in the fluid rest frame.

In \S\ref{sec:scalareom} we review how the scalar field equations of motion set the
rest frame sound speed and require non-adiabatic stress.   
The separate universe construction 
for long wavelength modes with initial adiabatic or curvature fluctuations  differs
above and below the sound horizon as shown in \S\ref{sec:scalarcurvature}.  In \S\ref{sec:scalariso}, we examine how  this construction
changes if the dark energy also has initial isocurvature fluctuations.

\subsection{Sound Horizon and Equations of Motion}
\label{sec:scalareom}

To close the equations of motion  (\ref{eqn:continuity}) of dynamical dark energy
 in general, we need to specify the
relationship between its pressure and energy density fluctuations, i.e.\ the sound speed.  In the scalar field model this is provided by the field equation 
given a specific Lagrangian.   More generally, for a dark energy component that 
accelerates the expansion $w_\de <-1/3$ and so typically the adiabatic sound speed\begin{equation}
c_{\de a}^2 \equiv \frac{\bar p_\de'}{\bar \rho_\de'} < 0.
\end{equation}
For the Navier-Stokes or Euler equation (\ref{eqn:continuity}) to be stable
\begin{equation}
c_{\de a}^2 \ne \frac{\delta p_\de}{\delta \rho_\de} \equiv c_{\de s}^2 > 0,
\end{equation}
where $s$ denotes synchronous gauge.  Thus an internal non-adiabatic stress $\Gamma_\de$ is required for dynamical dark energy  \cite{Hu:1998kj}.

In terms of the scalar field, internal non-adiabatic stress arises from the separate kinetic and potential contributions
to the energy density and pressure.  In that case the local energy density does not uniquely
specify the local pressure  as it is 
 possible to specify a sound speed that is independent of
$w_\de(a)$.  For definiteness, we take the Lagrangian
\begin{equation}
P(X,Q) = \Lambda_X \left( \frac{X}{ \Lambda_X} \right)^n - V(Q)
\end{equation}
as an example. 

  The sound speed for the fluid is best defined in the rest frame ``$r$"
or equivalently the constant field gauge \cite{ArmendarizPicon:2000dh}
\begin{eqnarray}
c_\de^2 &\equiv&   \frac{\delta p_{\de r}}{\delta \rho_{\de r}  } = \frac{P_{,X}}{2 P_{,XX} X + P_{,X}} = \frac{1}{2 n-1} ,
\end{eqnarray}
where
\begin{align}
\delta p_{\de r} &= \delta p_\de -  \bar p_\de' \frac{ a H}{k} \frac{u_\de}{1+w_\de}, \nonumber\\
\delta \rho_{\de r} &= \delta \rho_\de - \bar \rho_\de'  \frac{ a H}{k} \frac{u_\de}{1+w_\de}.
\end{align}
The sound horizon of this system is then defined by the wavenumber $k_\de$ where
\begin{equation}
\frac{c_\de k_\de }{ aH }=1.
\end{equation}
The pressure fluctuation therefore carries internal non-adiabatic    stress
 \begin{equation}
 p_Q\Gamma_Q =   \delta p_Q - c_{Qa}^2 \delta \rho_Q,
 \end{equation}
which we shall see also generates an entropy fluctuation $S_{\de \mr}$ dynamically.

 For the two component system, the equations of motion for the synchronous gauge perturbations
 (\ref{eqn:hL}) and (\ref{eqn:continuity})
 become
 \begin{align}
\delta_\de' +3(c_{\de s}^2 - w_\de) \delta_\de &= -\frac{k}{a H} u_\de + (1+w_\de) \delta_\br' ,
\label{eqn:QmsystemC}\\
u_\de' + (1-3 w_\de) u_\de &= \frac{k}{a H} c_{\de s}^2 \delta_\de, 
\label{eqn:QmsystemE}
\end{align}
for the scalar field and
\begin{align}
\delta_\br'' + \left(2 + \frac{H'}{H} \right) \delta_\br' &= \frac{3}{2}\frac{H_0^2}{H^2} \sum_J \Omega_J F_J (1+ 3 c_{J s}^2 ) \delta_J , 
\label{eqn:QmsystemD}
\end{align}
for the matter,
where $\delta_{\mr}=\delta_\br$, $c_{\mr s}^2=0$, $F_{\mr}=a^{-3}$ and
\begin{equation}
c_{\de s}^2 \delta_\de   = c_\de^2\delta_\de  +3(c_\de^2 - c_{\de a}^2) \frac{ a H }{k }u_\de.
\end{equation}
To specify the background evolution,  we can either fix $w_\de$ by hand and leave
the corresponding potential implicit or solve the $\de$ background equations of motion\begin{equation}
\bar Q'' + \frac{3}{2}(2 c_\de^2 -1 - w_T) \bar Q' + \frac{c_\de^2}{H^2} \frac{V_{,Q}}{P_{,X}}=0,
\label{eqn:Qfield}
\end{equation}
for a given potential $V$.  
Here
the total equation of state $w_T=p_\de/(\rho_\mr+\rho_\de)$ for the two component 
system and we construct
\begin{equation}
w_\de =\frac{\bar p_\de}{\bar \rho_\de}= \frac{P}{2 P_{,X} X - P} \Big|_{\bar Q}
\end{equation}
from the background solution $\bar Q$.

  Below the sound horizon, $\de$ density fluctuations are pressure supported and become
negligible compared with  the dark matter.   Above the sound horizon, dark energy fluctuations influence the growth of matter density perturbations and thus the separate universe construction.   Above the total Jeans scale, the separate universe construction provides
a \trueword\ separate universe.    The correspondence between the total Jeans scale and
the sound horizon depends on the initial conditions as we shall see next.
 
 %==================== Figure ====================
\begin{figure}
\center
\includegraphics[width = 0.45\textwidth]{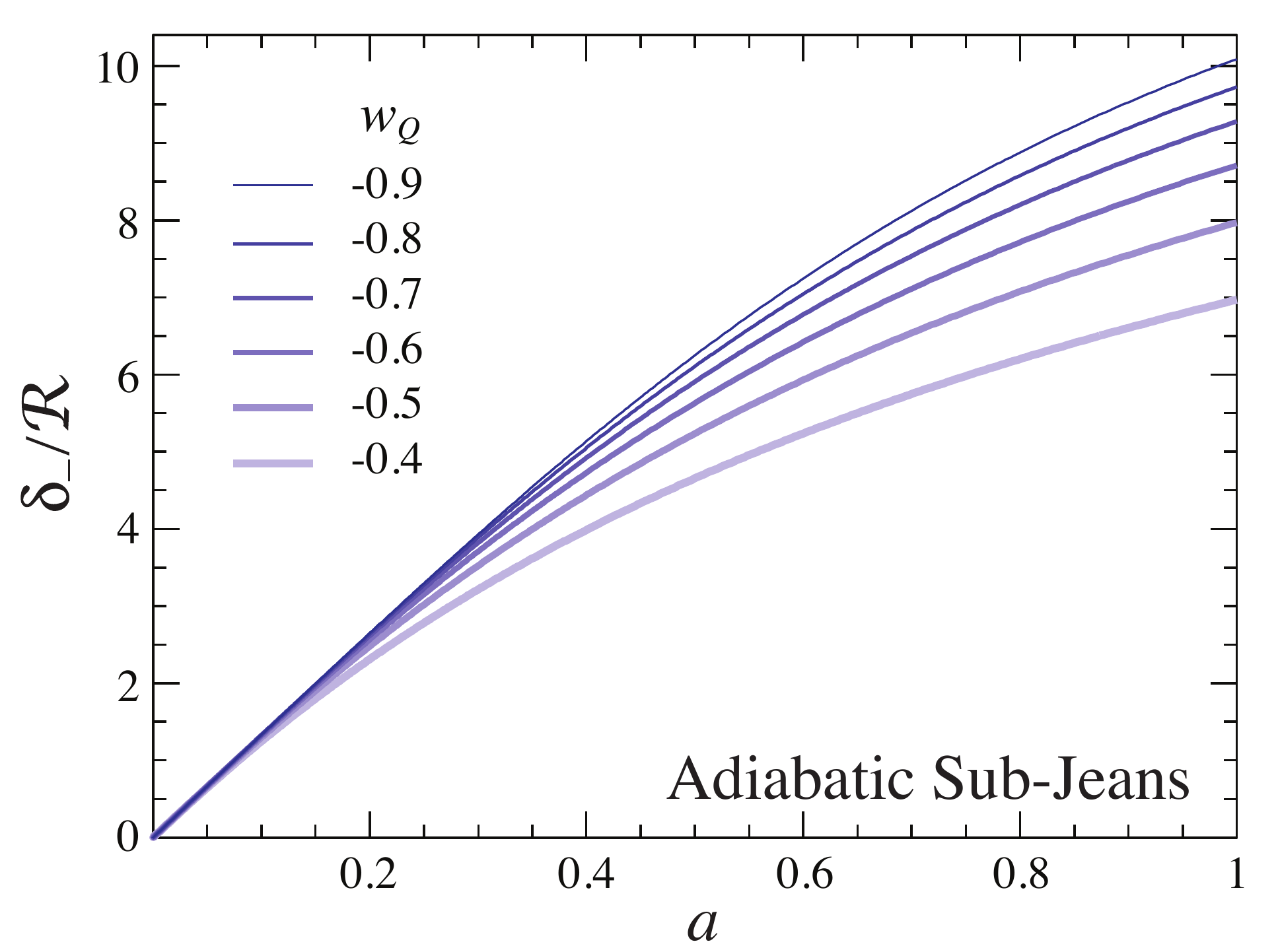}
\caption{Adiabatic matter density perturbation evolution $\delta_\br$  relative
to the initial curvature fluctuation ${\cal R}$ for an arbitrary wavenumber on sub Jeans scales.  Pressure support in the scalar field $\de$ slows the growth of fluctuations as $w_\de$ increases and dark energy domination occurs earlier. Here
  and throughout the figures, we
take $\Omega_\mr=1-\Omega_\de=0.3$. }
\label{fig:delta}
\end{figure}
%==================== Figure ====================

%==================== Figure ====================
\begin{figure}
\center
\includegraphics[width = 0.45\textwidth]{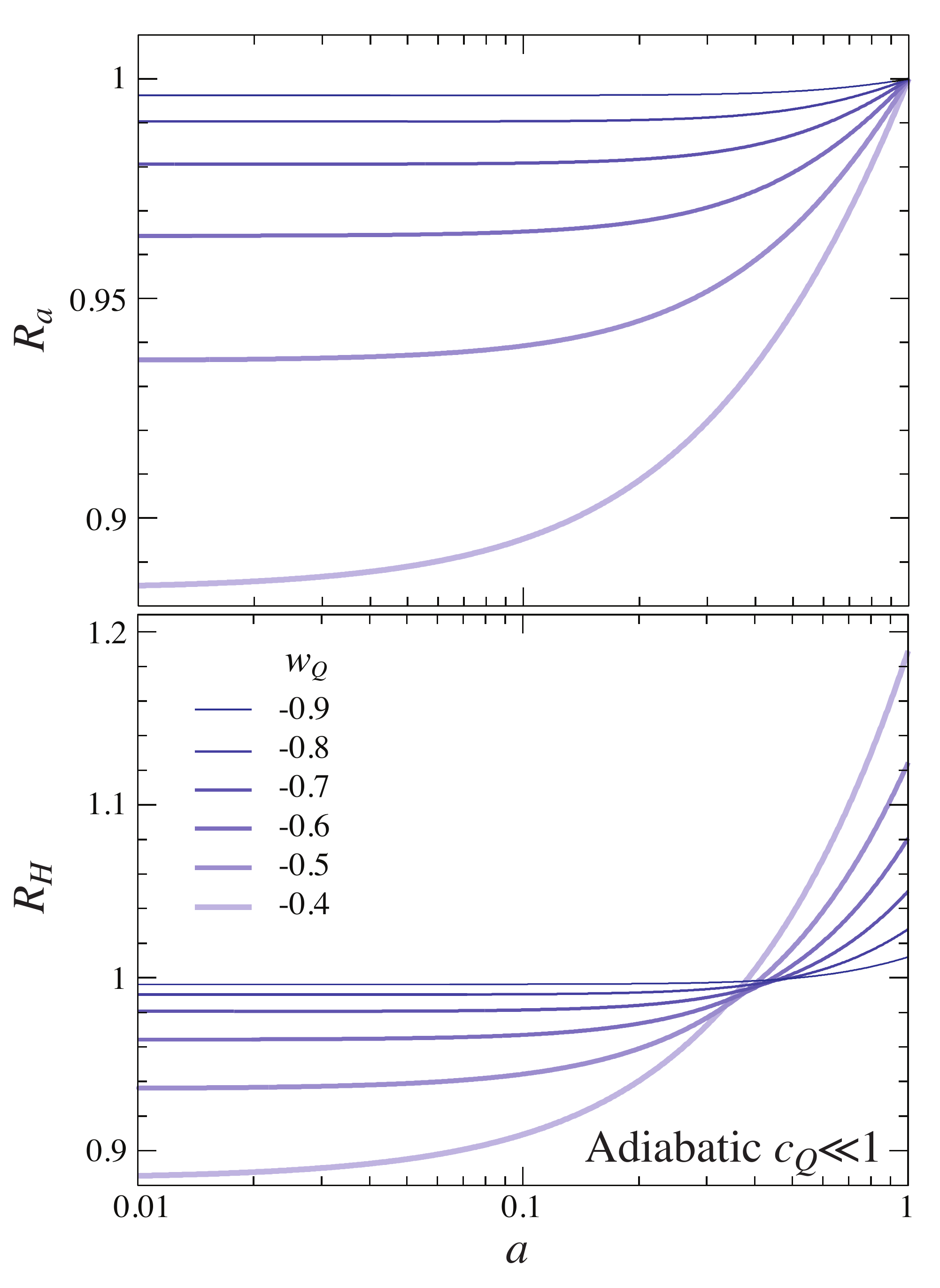}
\caption{Ratios of super to sub Jeans scale  separate universe changes in the scale factor $\delta \ln a$ (top) and Hubble
rate $\delta \ln H$ (bottom) for the same long wavelength
density $\delta_0$ today.  
Above the Jeans scale,  the scale factor is always closer to global  whereas the Hubble rate 
is closer at early times and further at late times. Observable responses to $\delta_0$ therefore
depend on its scale, but as $w_\de \rightarrow -1$ this difference vanishes.  Here
initial conditions are adiabatic and $c_\de \ll 1$. }
\label{fig:curvature}
\end{figure}
%==================== Figure ====================

\subsection{Initial Curvature Perturbations}
 \label{sec:scalarcurvature}

For the usual case of adiabatic perturbations that originate from initial comoving curvature fluctuations ${\cal R}$ in the matter dominated epoch, the matter density fluctuation growth takes on a scale-free form
when the mode is either well
above or below the $\de$ 
sound horizon.   In this case the sound horizon defined in constant field gauge corresponds
to the total Jeans scale during the acceleration epoch  \cite{Creminelli:2009mu}.

For simplicity let us assume that $P(X,Q)$ has been constructed so that in 
the global background $0> w_\de=c_{\de a}^2=$\, const.   Then at the initial epoch $a_i$ the universe
is matter dominated and the growing mode of adiabatic or initial curvature fluctuations ${\cal R}$
is 
\begin{equation}
\delta(a_i) = \frac{2}{5} \left( \frac{k}{a_i H_i} \right)^2 {\cal R} \propto a_i.
\end{equation}
Inspecting the equations of motion (\ref{eqn:QmsystemC}), (\ref{eqn:QmsystemE}) and (\ref{eqn:QmsystemD}), 
we obtain
\begin{eqnarray}
\delta_\de(a_i) &=& \frac{(5-6 c_\de^2)(1+ w_\de)}{ 5 + 9 c_\de^2 - 15 w_\de} \delta(a_i) ,\nonumber\\
u_\de(a_i) &=& \frac{k}{a_i H_i} \frac{2 c_\de^2 (1+ w_\de)}{ 5 + 9 c_\de^2 - 15 w_\de}  \delta(a_i).  
\label{eqn:curvatureIC} 
\end{eqnarray}
Note that there is an  entropy fluctuation induced by the curvature fluctuations
\begin{equation}
S_{\de \mr} = \frac{15 w_\de - 15 c_\de^2}{ 5 + 9 c_\de^2 - 15 w_\de } \delta_\br(a_i)
\end{equation}
and the internal non-adiabatic stress.   Here $S_{\de\mr} \propto \delta_\br$ and so $|S_{\de \mr}| 
\ll |{\cal R}| \sim |\eta_T|$ outside the horizon when $k/aH \ll 1$ unlike the isocurvature conditions discussed
in the next section.  Equivalently, the entropy fluctuation vanishes as $a_i \rightarrow 0$
and the background dark energy density in the separate and global universe are initially the same. 

%==================== Figure ====================
\begin{figure}
\center
\includegraphics[width = 0.45\textwidth]{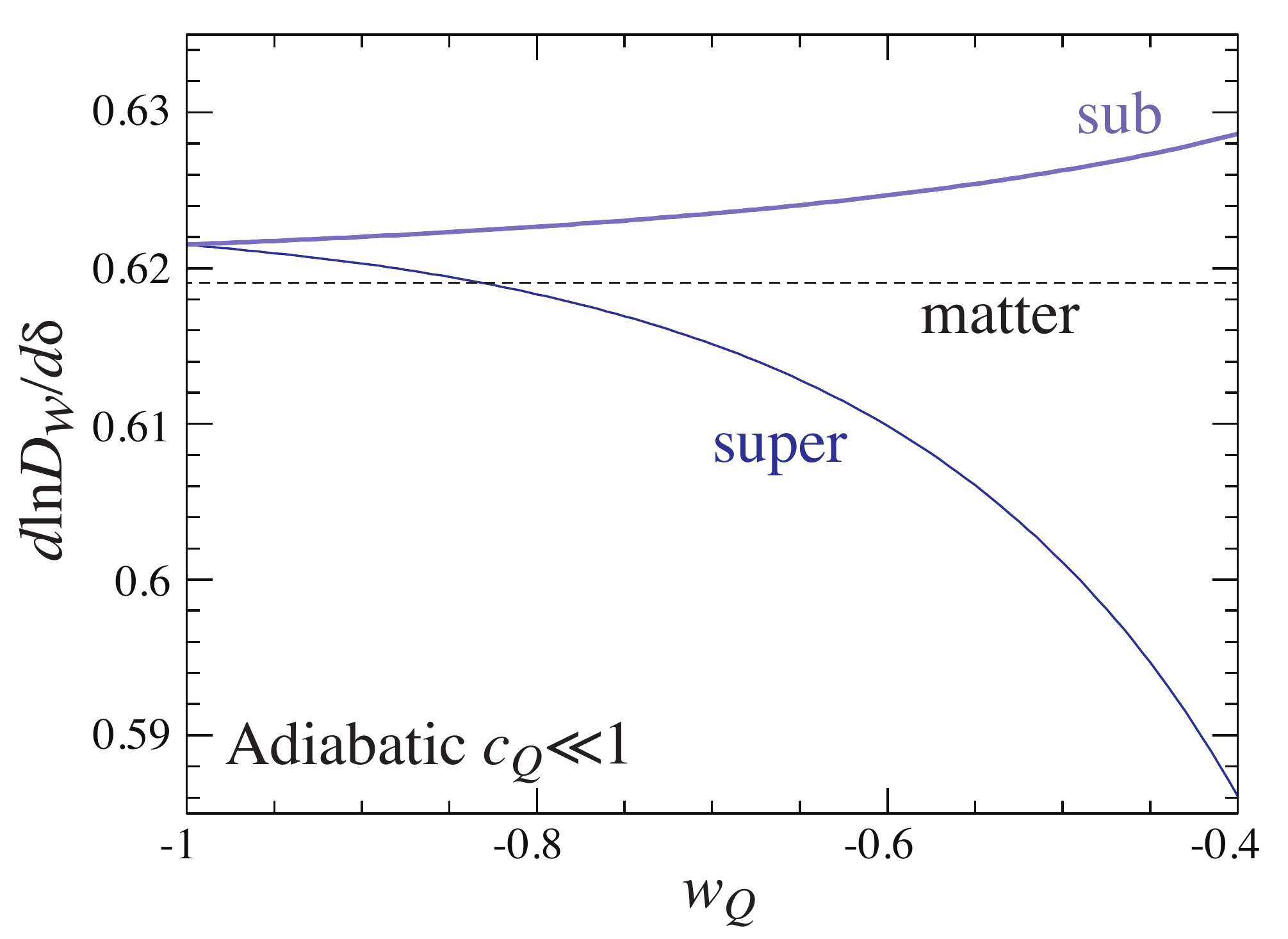}
\caption{Response of the short wavelength 
linear growth function $D_W$ to the long wavelength density field $\delta$ at $a=1$ for
adiabatic initial conditions and scalar field dark energy with $c_\de \ll 1$.    This response
depends on whether the long wavelength is super or sub Jeans scale.  Long wavelength scale dependence of the
squeezed bispectrum and trispectrum would result but vanishes if $w_\de \rightarrow -1$.   For
reference, the second order response in the matter dominated regime $13/21$ is also shown (dashed).
 }
\label{fig:growresponse}
\end{figure}
%==================== Figure ====================

For the sub Jeans case of $k\gg k_\de$, the dark energy perturbations are negligible
compared with the matter and the system reduces to the familiar case $\delta_\br\approx\delta_-$
\begin{equation}
\delta_- '' + \left(2 + \frac{H'}{H} \right) \delta_- ' = \frac{3}{2}\frac{H_0^2}{H^2}  \frac{\Omega_\mr}{a^{3}} 
\delta_- .
\end{equation}
We show this growth function in Fig.~\ref{fig:delta} for various $w_\de$.   As $w_\de$ increases,
dark energy domination occurs earlier for the same $\Omega_\mr$ and pressure support in $\de$
has a larger impact on the matter growth.

For the super Jeans case of $k\ll k_\de$, the growth of $\delta_\br$ in the dark energy dominated epoch depends on the
sound speed if relativistic $c_\de \sim 1$.   In this case the Jeans scale is near the horizon.
The super and sub Jeans scale differences are difficult to measure and also involve
relativistic effects in relating synchronous gauge quantities to direct observables.

It is therefore interesting to consider the $c_\de \ll 1$ limit where we solve
\begin{eqnarray}
\delta_\de' -3 w_\de  \delta_\de &=& (1+w_\de) \delta_+' , \\
\delta_+ '' + \left(2 + \frac{H'}{H} \right) \delta_+ ' &=& \frac{3}{2}\frac{H_0^2}{H^2} \left[  \frac{\Omega_\mr}{a^{3}} \delta_+
+\Omega_\de F_\de \delta_\de \right] ,\nonumber
 \end{eqnarray}
 and take $\delta_\br \approx \delta_+$.   
 Notice that in this limit the Euler equation (\ref{eqn:QmsystemE}) for $u_\de$ has negligible source
 from gradients in $\delta_\de$  and with the initial conditions (\ref{eqn:curvatureIC}) its value 
 remains negligible.   Thus we have a \trueword\ separate universe above the
 sound horizon which plays the role of the Jeans scale for initial curvature fluctuations \cite{Creminelli:2009mu}.
 
 This sets the change in the separate universe
 scale factor $\delta \ln a=  \ln a_W -\ln a$.   In Fig.~\ref{fig:curvature} (top), we show the
 ratio of this change above to below the Jeans scale for
 the same value of $\delta(a=1)=\delta_0$
 \begin{equation}
R_a \equiv\frac{\delta \ln a_+}{\delta \ln a_-} = \frac{ \delta_+(a)}{\delta_-(a)} \frac{\delta_{-}(1)}{\delta_+(1)}.
\end{equation}
For definiteness we take $\Omega_\mr=1-\Omega_\de=0.3$ here and throughout the examples.
 Since above the Jeans scale, $\delta_\br$ grows more
relative to its initial value, it is actually smaller in value at $a<1$ once normalized to today.  This difference goes to zero as $w_\de\rightarrow -1$ since
the gravitational source to $\delta_\de$ in Eq.~(\ref{eqn:QmsystemC}),  $(1+w_Q)\delta' \rightarrow 0$.

The difference in $a_W(a)$ between the super and sub Jeans scale separate universes 
implies that the Hubble rate also differs.  Using Eq.~(\ref{eqn:dh2}), for the same
value of $\delta_0$ the ratio is
\begin{equation}
R_H \equiv \frac{\delta \ln H_+}{\delta \ln H_-} = \frac{ \delta_+'(a)}{\delta_-'(a)} \frac{\delta_{-}(1)}{\delta_+(1)}.
\end{equation}
In the Hubble rate, the super Jeans scale separate universe is closer to the global universe that the sub Jeans scale one at early times and farther at late times as the universe begins to accelerate.  
The latter reflects the enhanced growth rate above the Jeans scale required to produce the
same $\delta_0$ today.

An interesting consequence of this behavior is that 
objects formed at high redshift will differ in their response to a long wavelength mode than those formed at low redshift.   We will address the 
implications for halo bias in future work.

This dependence on scale and redshift also changes the response of the linear growth  
$d\ln D_W/d\ln \delta$ of Eq.~(\ref{eqn:DWresponse})  above and below the Jeans scale.
This is shown in Fig.~\ref{fig:growresponse}.   Since relative to the same $\delta_\br$ today, its
amplitude at high redshift is smaller above the Jeans scale as shown in Fig.~\ref{fig:curvature}, the linear growth response is also smaller.   
In principle this would lead to an observable change in the matter bispectrum, trispectrum and
super sample power spectrum covariance but 
the size of this change is small for observationally viable values of $w_\de$.

%==================== Figure ====================
\begin{figure}
\center
\includegraphics[width = 0.45\textwidth]{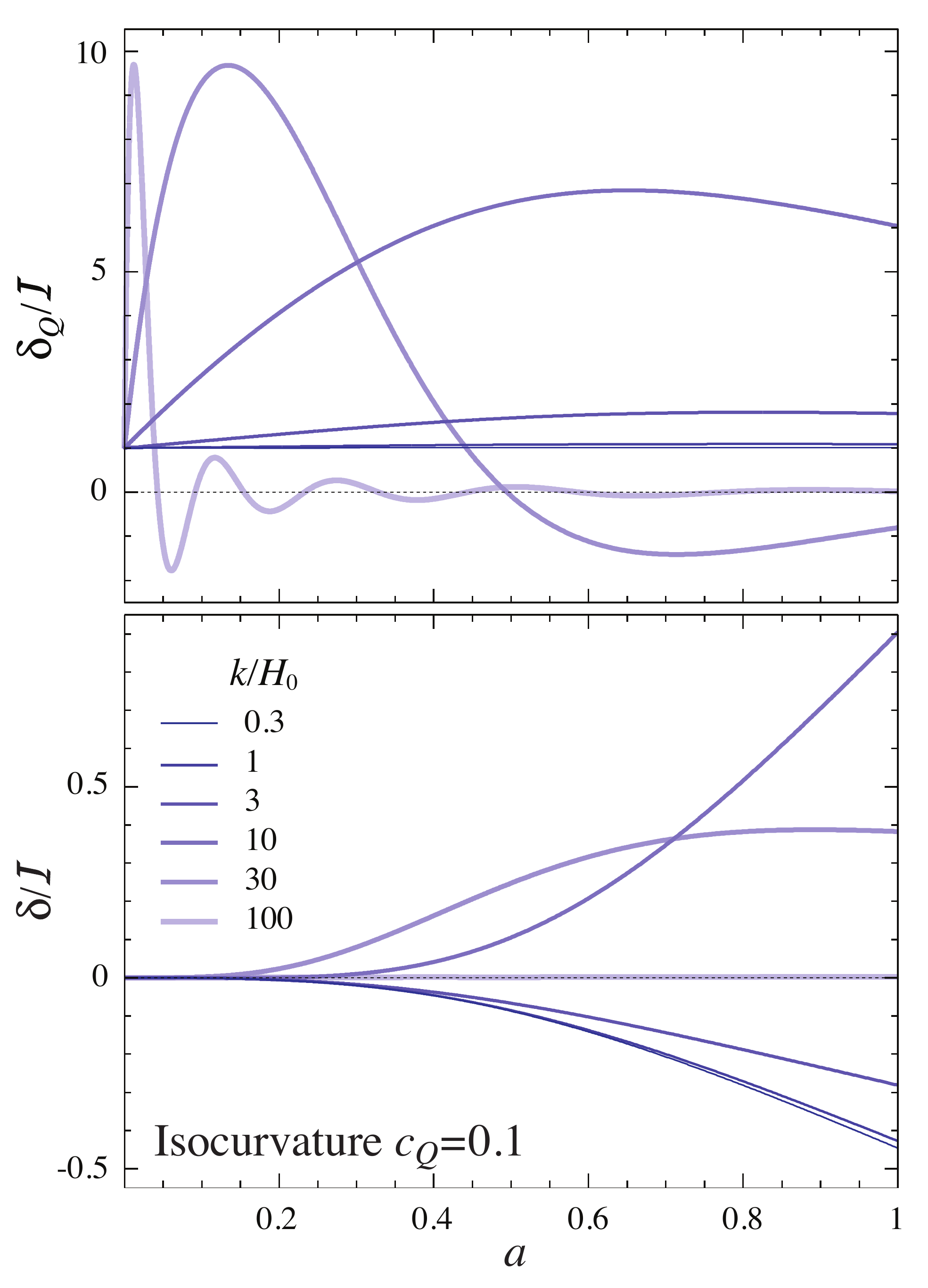}
\caption{Growth of the dark energy and matter density perturbations $\delta_\de$ and $\delta_\br$ from dark energy isocurvature initial conditions ${\cal I}$ for $c_\de=0.1$.
Dark energy perturbations (top) are frozen outside the horizon $k/H_0 <1$, grow between
the horizon and sound horizon, and oscillate below the sound horizon.   Matter perturbations (bottom)
are anticorrelated with ${\cal I}$ outside the horizon, grow in correlation  with it between the horizon and sound horizon, and are independent of it below the sound horizon.   By adding
isocurvature modes to the adiabatic modes, the separate universe construction and observable response can depend on scale even for $w_\de \approx -1$.
 }
\label{fig:isocurvature}
\end{figure}
%==================== Figure ====================

\subsection{Initial Dark Energy Perturbations}
\label{sec:scalariso}

For pure curvature initial conditions, we have seen that the difference between
super and sub Jeans scale growth and hence the scale dependencies of separate
universe responses of observables vanish as $w_\de \rightarrow -1$.  This is because the gravitational source to these dark energy density
fluctuations vanishes in this limit.   The difference can be much larger if instead these fluctuations were
provided by the initial conditions.   

If the field fluctuations associated with these initial conditions were nearly frozen outside the horizon, then large scale fluctuations would
survive to the current epoch  \cite{Gordon:2004ez}.  
In the field equation (\ref{eqn:Qfield}), this occurs when the slope of the potential $V_{,\de}$ is too small
to overcome the Hubble drag.   The field then only rolls by a small amount during its
cosmic evolution.    In that case  $V_{,\de} \approx $const for any smooth potential, and the field equation becomes
a Bernoulli equation for $Q'$ which does not depend on the field value itself.    
By making $V_{,\de}$ arbitrarily small compared with $V$,
we can bring the expansion history as close to $\Lambda$CDM as desired.   We therefore consider for simplicity 
the limiting case where
 \begin{eqnarray}
w_T(a) %&=& \frac{\bar p_\de}{\bar \rho_\mr + \bar \rho_\de} \nonumber\\
&=&- \frac{\Omega_\de}{\Omega_\mr a^{-3} + \Omega_\de},
\end{eqnarray}
in a flat $\Omega_\mr+\Omega_\de=1$
universe.

In this case, $Q'$ reaches a terminal velocity independent of its initial value and
the adiabatic sound speed becomes
\begin{equation}
c_{\de a}^2 =\frac{ x(c_\de^2-x^2) \sqrt{1+x^2} - c_\de^2(1+x^2) \sinh^{-1}x}
{(1+ x^2) ( x\sqrt{1+x^2} - \sinh^{-1}x)},
\end{equation}
where 
\begin{equation}
x = a^{3/2}\sqrt{\frac{\Omega_\de}{\Omega_m}}
\end{equation}
parameterizes the transition between matter and dark energy domination.

Note that in the matter dominated epoch $x\ll 1$ and 
\begin{equation}
c_{\de a}^2 = -\frac{c_{\de}^2 +3}{2}.
\end{equation}
Since the adiabatic sound speed $c_{\de a}^2 = p_\de'/\rho_\de'$, the small 
evolution of the energy density and pressure makes $c_{\de a}^2 \ne w_\de \approx -1$.

Inspecting the equations of motion, we find that initial conditions in the matter dominated epoch are
\begin{eqnarray}
\delta_{\de}(a_i) & =& {\cal I}, \nonumber\\
u_\de(a_i) &=& -\frac{2}{9}\frac{k}{a_i H_i} {\cal I}, \nonumber\\
\delta_\br(a_i) &=& -\frac{1}{3}\frac{\Omega_\de}{\Omega_\mr} a_i^3 {\cal I} .
\label{eqn:isoIC}
\end{eqnarray}
The constant ${\cal I}$ is equivalent to an initial entropy fluctuation ${\cal I}= (1+w_\de) S_{\de \mr}$, but
given that $w_\de\approx -1$,  this notation is more convenient.
Note that 
\begin{equation}
\eta_T(a_i) = -\frac{1}{9} \frac{\Omega_\de}{\Omega_\mr} a_i^3  {\cal I} ,
\end{equation}
reflecting the isocurvature initial conditions $|\eta_T| \ll |{\cal I}|$.   These quantities should be added to
those generated by the initial curvature fluctuations ${\cal R}$ and so their relative strengths
are determined by ${\cal I}/{\cal R}$ and their correlation.

In Fig.~\ref{fig:isocurvature} (top) we show the evolution of $\delta_\de$ from these isocurvature conditions as 
a function of scale for an example with $c_\de=0.1$.  
On scales larger than the current horizon $k/H_0$, the density perturbations are frozen even through the
acceleration epoch as expected.    We can analytically verify this behavior by dropping the velocity divergence
source $ku_\de/aH$ in the $\de$ continuity equation (\ref{eqn:QmsystemC}).   The $\de$ system is then
solved by 
\begin{eqnarray}
\delta_\de(a) & =& {\cal I} ,\nonumber\\
u_\de(a) &=& -\frac{1+c_\de^2}{3 (c_\de^2- c_{\de a}^2)}\frac{k}{aH} {\cal I},  \qquad \frac{k}{aH} \ll 1.
\end{eqnarray}
These dark energy perturbations induce a growing mode in the matter fluctuations of opposite sign (see Fig.~\ref{fig:isocurvature}, bottom).  
In terms of the separate universe, a positive change in the dark energy density acts like a universe with 
a larger cosmological constant.  In the construction of Eq.~(\ref{eqn:fstrue}) we introduce an entropy component
with
\begin{equation}
\Omega_\fake F_\fake(a) = \Omega_\de(1+w_\de) S_{\de \mr} = \Omega_\de {\cal I}
\end{equation}
and this represents a \trueword\ separate universe with a constant change to the cosmological constant
as expected.  Technically, the curvature fluctuation is changing at the Hubble rate during matter domination and
growing logarithmically during dark energy domination but its value is suppressed by $(k/aH)^2$ and 
has negligible impact since initial curvature fluctuations must also exist.

 Between
the horizon and sound horizon, the dark energy density fluctuation grows due to the non-gravitational velocity divergence
in its continuity equation.   Unlike in the case of curvature fluctuations, the total Jeans scale in the
acceleration epoch is the horizon and not the sound horizon.   
In the Euler equation (\ref{eqn:QmsystemE}), even though
density gradients from $\delta_\de$ do not source non-gravitational flows for $c_\de \ll 1$, their initial values are set by 
the isocurvature conditions (\ref{eqn:isoIC}) and they grow until horizon crossing.   At this point, they become
large enough to cause  a violation of the  \trueword\ separate universe.  The divergence of the flow then impacts $\delta_\de$ through
the continuity equation (\ref{eqn:QmsystemC}) which sources $\delta_\br$ through
Eq.~(\ref{eqn:QmsystemD}) causing it to change signs.   In this regime a positive initial dark energy
density fluctuation leads to a positive and growing matter fluctuation.   

Below the sound horizon, the dark energy density fluctuation oscillates and decays.  The matter fluctuations then
are also suppressed and the separate and global universe coincide in parameters.   
In principle a small $c_{\de}\ll 1$ allows these effects to produce novel separate universe
responses between the horizon and sound horizon.    However, the matter density perturbations associated
with the curvature fluctuations ${\cal R}$ also grow during this regime for $\Omega_\mr \sim 0.3$ where the
universe is not completely dark energy dominated.   One must therefore arrange the initial spectrum of
${\cal I}$ to produce a sizable change in $\delta_\br$ of the desired wavelength at the current epoch.   Finding an early universe mechanism
to generate such dark energy isocurvature fluctuations is beyond the scope of this work.

\section{Discussion}
\label{sec:discussion}

In this work, we have shown how to construct a separate universe to absorb the entire 
growth history of a long wavelength
density perturbation of a multicomponent system into the cosmological background from the perspective
of the non-relativistic matter.
 By exactly matching the acceleration equation to the
synchronous gauge matter density fluctuations, we  extend the validity of the approach to 
scales smaller than the Jeans length where non-gravitational effects play a role.   
Above the  Jeans scale, the construction also satisfies the Friedmann equation with
real energy densities and a curvature that is constant in comoving coordinates.   Below
the Jeans scales, the curvature evolves and in the separate universe Friedmann equation
acts like a fake density component.   In both cases, the matter evolution on small scales
is correctly modeled.   

Once the long wavelength density fluctuation is absorbed into the background, we can assess its
impact on  small scale cosmological observables as a change in the expansion rate or the cosmological parameters that drive it.
Our construction highlights the fact that its influence is nonlocal in time.   For the same long wavelength density fluctuation, its impact on small scale observables at the same epoch depends
on its entire growth history.    If this growth depends on scale as in the case of the
super and sub Jeans scale fluctuations, then the response also
becomes dependent on the scale of the long wavelength mode.

As a concrete illustration, scalar field dark energy with a finite sound speed introduces its sound horizon
to the Jeans scale
of the system.   For the same long wavelength density perturbation today, the different
growth histories imply different separate universes and hence different responses in short
wavelength observables.   In particular, we have highlighted the scale dependent response
to the linear growth rate for adiabatic fluctuations and the novel changes that can occur
if initial dark energy isocurvature perturbations are also present.     

By employing cosmological simulations of the separate universe,
this technique should prove useful for studying the analogous scale dependent responses
in the nonlinear matter and halo power spectrum, super-sample covariance, bispectrum,
trispectrum and halo abundance and bias.    Likewise
other systems such as massive neutrino and modified gravity models possess scale dependent
long-wavelength growth that can also be studied with
these methods.   We leave these topics for future work.

\acknowledgments{%%%%%%%%%%%%%%%%%%%%%%%%%%%%%%  
We thank Uros Seljak and Masahiro Takada  for useful discussions.  
WH was supported by U.S.~Dept.\ of Energy
contract DE-FG02-13ER41958,  NASA ATP NNX15AK22G, and 
the Kavli Institute for Cosmological Physics at the University of
Chicago through grants NSF PHY-0114422 and NSF PHY-0551142 and an endowment from 
the Kavli Foundation and its founder Fred Kavli.  
CC and ML are supported by grant NSF PHY-1316617.}

\bibliography{faksu}

\end{document}